# The geodynamo for non-geophysicists

Tobias Vorbach und Friedrich Herrmann

Abteilung für Didaktik der Physik, Karlsruhe Institute of Technology, D-76128 Karlsruhe, Germany

The geodynamo usually appears as a somewhat intimidating subject. Its understanding seems to require the intricate theory of magnetohydrodynamics. The solution of the corresponding equations can only be achieved numerically. It seems to be a subject for the specialist.

We show that one can understand the basics of the functioning of the geodynamo solely by using the well-known laws of electrodynamics.

The topic is not only important for geophysicists. The same physics is the cause for the magnetic fields of sun-like stars, of the very strong fields of neutron stars, and also of the cosmic magnetic fields.

## I. INTRODUCTION

In the physics lecture, the origin of the magnetic field of the Earth is usually only briefly mentioned. Roughly, the students learn the following: The field does not originate from a permanent magnet, because the temperature inside the Earth is too high. It is caused by electric currents which are generated in the same way as with a self-excited dynamo: by the movement of an electrical conductor in a magnetic field. This "geodynamo" is driven by the thermal convection of the liquid iron inside the Earth.

To say more, one might argue, is not possible in the scope of the physics lecture. A more detailed treatment would require the theory of magnetohydrodynamics and mainly consist of computing. Textbooks on the Earth's magnetic field indicate that the subject is extremely complex[1].

We would like to confront such an attitude or habit with another view. We think that physics students should learn more about the geodynamo for the following reasons:

- The Earth's magnetic field belongs to a species that is widespread in the cosmos: Not only the Earth and other planets but also stars and galaxies have magnetic fields. Some of them have such great field strengths that they eclipse even the strongest laboratory fields. For example, the magnetic field of a neutron star is so intense that one liter of it can weigh 1 kg. The way they are generated is essentially the same as that of the Earth's magnetic field.

- Just as we treat the basic laws of meteorology in the thermodynamics and mechanics lecture, but leave the concrete calculation of weather and climate to the meteorologists, we should also address the basics of the magnetic field inside the Earth in the physics lecture and leave its calculation to the geophysicists.

As long as we restrict ourselves to the few above-mentioned remarks about the Earth's magnetic field, the students may ask themselves some important questions, such as

- With the self-excited dynamo, the path of the electric current through the conductors is well-defined, as is the movement of the conductors. It seems incredible that a dynamo effect takes place when the electric current is allowed to flow "as it likes" and the conductors can move "as they like".

- We know that the technical dynamo has to be started. Who has started the geodynamo?

- Why does the dynamo effect exist in the Earth (and other celestial bodies), but not in systems of our everyday environment?

We will discuss and answer these questions.

Our description of how the geodynamo works corresponds to about one teaching hour of the physics lecture.

In section II we will describe the phenomenon: the structure of the Earth and the structure of the magnetic field. Section III explains how the geodynamo is working. However, it will not yet be clear under which circumstances the geodynamo effect actually occurs. This question is clarified in section IV. Finally, it is still to be discussed how the dynamo is started. This happens in section V. In section VI the results are summarized.

## II. THE STRUCTURE OF THE EARTH AND OF ITS MAGNETIC FIELD

### A. The Earth

Fig. 1 shows a cross-section of the Earth. The core, which has a radius of 3480 km, consists mainly of iron and is electrically conductive. (The conductivity is about 1/10 of that of iron under normal conditions.) The mantle consists of rock whose electrical conductivity is about 5 orders of magnitude smaller. We can consider it an insulator. The inner core with a radius of 1210 km is solid, the outer core is liquid. The viscosity of the iron in the outer core is not very different from that of liquid iron on the Earth's surface. The viscosity of the mantle, on the other hand, is so high that we can consider it solid. The temperature in the core and mantle decreases from the inside to the outside. In the center of the Earth, it is about 6000 K, at the surface of the core it is slightly more than 4000 K. Since a heat transport takes place from inside to outside, the Earth cools down slowly. This cooling process, however, has a duration comparable to the age of the universe. We can, therefore, assume a temperature gradient from the inside to the outside that is constant in time.

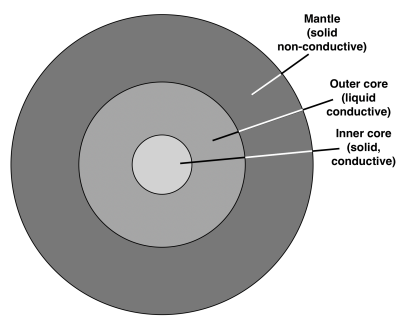

Fig. 1. The structure of the earth. The outer core meets the conditions for the dynamo effect to occur: it is liquid and electrically conductive.

In the region we are interested in, namely the outer core the heat is transported by convection: The liquid metal moves at a speed of a few km/a. (Compared to other "geological" velocities, such as those of the continental plates, this velocity is quite high.) However, this convection current is not simply up and down. Due to the rotation of the Earth, any translational motion is overlaid by a rotational motion, a phenomenon that we know from the air currents in the atmosphere. Altogether we have a screw-shaped motion.

### B. The field

Where the field can be easily observed, i.e. outside of the Earth, it is approximately a dipole field, similar to the field of a bar magnet or a cylindrical coil. But inside of the Earth, the spatial distribution of the field strength is far from being that of a dipole field. Fig. 2 shows calculated field lines[2]. It has an irregular and intertwined structure with variations on a length scale of about 100 km.

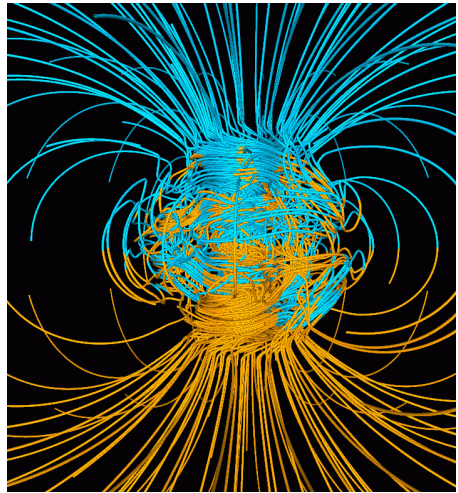

Fig. 2. Calculated magnetic field[2]. In the earth's core it is irregular and intertwined.

**C. The temporal development**

If one examines the field in a time interval that is short according to human time scales, one gets the impression that the field is constant in time. However, when looking at it over several hundred thousand years the time development of the field can be seen[2].

## III. THE MECHANISM OF GENERATING THE MAGNETIC FIELD

**A. The technical self-excited generator**

The Geodynamo basically works like a self-excited technical generator. Let us briefly recall how such a generator works. The stator is an electromagnet that creates the excitation field. A current is induced in the rotor coil, which rotates in the stator field. This same current is passed through the stator coil.

The self-excited dynamo has some characteristics that we also find in the geodynamo, and which we would like to highlight here.

- In order for the self-excited dynamo to start, the stator must first be supplied from another source, because as long as no current flows in the stator coils, no current is induced in the rotor coils.
- If the dynamo runs too slowly, it can "extinguish".
- If the rotor runs sufficiently fast, an arbitrarily small magnetic field is sufficient to start the dynamo effect. The faster the dynamo runs, the more unstable becomes the currentless state.

**B. How the geodynamo doesn't work**

For a technical dynamo, the movement of the electrical conductors is well-defined, as is the path of the electrical current. The Earth achieves the same result, although the conductors seem to move without any plan, and although the electric current has no predetermined paths. In order to understand how it performs this miracle, let us first discuss a situation in which there is no dynamo effect, and understand the reason for it.

In the following, we will deal with different contributions to the total magnetic field. We call the corresponding field strengths $\mathbf{B}_0$, $\mathbf{B}_1$, and so on. Accordingly, we designate electric currents with $\mathbf{j}_0$, $\mathbf{j}_1$, … (the symbol of the current density) and movements with $\mathbf{v}_0$, $\mathbf{v}_1$, (the symbol of velocity).

To get a dynamo effect, we have to assume that there is a magnetic field at the beginning. Let's call it **B**$_0$. We will discuss in section 5 where this initial field may come from. For the moment it is just "fallen from the sky". We now hope to get a new magnetic field out of it by a suitable movement of an electrically conductive liquid, so that the induction process continues by itself even though we subsequently switch off the source of **B**$_0$.

We begin by moving our liquid within the initial field perpendicularly to the direction of the field vector **B**$_0$, Fig. 3. (Black arrows represent movements, the red tubular shapes represent the magnetic flux and blue arrows the electric current density vector.)

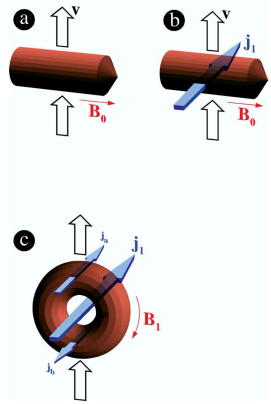

Fig. 3. A translational movement (a) in the magnetic field **B**$_0$ results in an electric current **j**$_1$ (b). The current **j**$_1$ causes the magnetic field **B**$_1$ (c). The movement of the liquid in **B**$_1$ gives rise to the new currents **j**$_a$ and **j**$_b$. As a result, the total current is shifted upwards. The liquid flow attempts to carry the electric current with it.

The induced electric current is perpendicular to **B**$_0$ and to the velocity according to the right-hand rule. The induced current itself causes a magnetic field **B**$_1$. What is the effect of the movement of the liquid in **B**$_1$? Above **j**$_1$ there is a current **j**$_a$, which flows in the same direction as **j**$_1$, and below is a current **j**$_b$, which flows in the opposite direction. The result of all three currents **j**$_1$, **j**$_a$ and **j**$_b$ together is an upwards shift of the total current. In other words: the liquid tries to drag the current **j**$_1$ with it. The greater the electrical conductivity, the better it succeeds in doing so. In the ideal case of a perfect conductor, this effect would be complete: The field would be "frozen" in the liquid.

But what happens when we now switch off **B**$_0$? The induced current is displaced a little by the liquid and at the same time, it decays. There is no dynamo effect. The reason: The movement was too simple.

**C. How the geodynamo works**

In fact, the dynamo effect can only arise if a second, independent movement is added to the first translational movement: a rotation around the direction of the translational velocity. The result is a helical motion. In the following, we explain how such a movement produces a magnetic field that has the same direction as the initial magnetic field. Figure 4 shows the process, broken down into 4 separate steps.

In each of them we consider only part of the problem: first only the translational component of the movement, then only the rotation, then again only the translation and finally once more the rotation. We also decompose the magnetic field and the electric current density vectors into components and consider only one of them at a time. This is permitted because the different components of the vector quantities velocity, magnetic field strength, and electrical current density add up linearly and therefore do not influence each other. Of course, we only get a small part of the total solution of the problem.

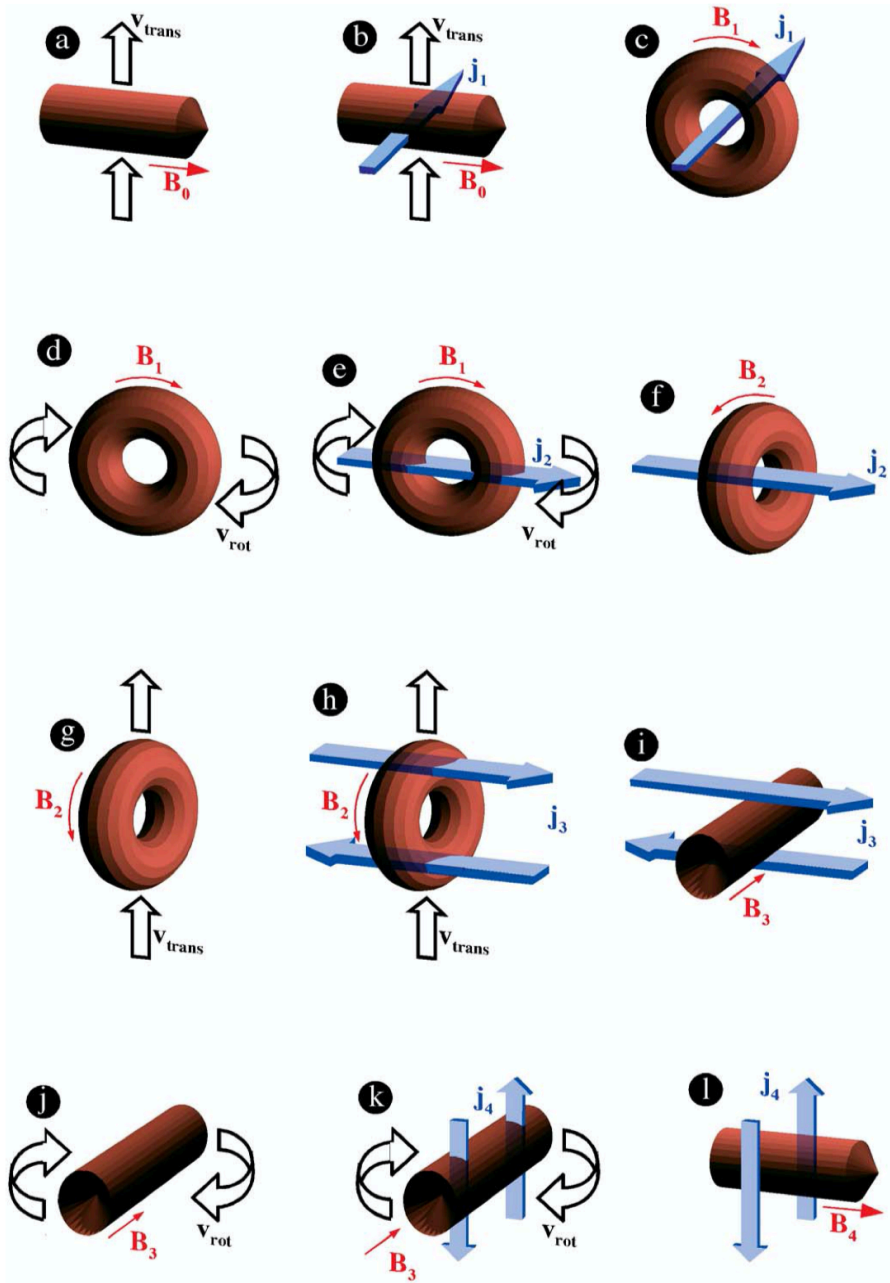

Fig. 4. Translation and rotation cause an initially existing field **B**$_0$ to be converted in several steps into a new field **B**$_4$, which has the same direction as **B**$_0$.

Figure 4a shows the initial magnetic field **B**$_0$ and the translational motion of the liquid. This movement within the magnetic field results in an electric current **j**$_1$, Figure 4b. For the sake of clarity, **B**$_0$ and the translational motion **v**$_{trans}$ are no longer shown in the next figure (4c). In the following, we are only interested in the effects of the current **j**$_1$. This current generates a magnetic field, Figure 4c. Figure 4d does no longer show **j**$_1$. Instead, we consider the effects of the rotational motion of the liquid within the field **B**$_1$. Figure 4e sketches the resulting electric current **j**$_2$. In Figure 4f, **B**$_1$ and **v**$_{rot}$ are no longer shown, but instead the magnetic field **B**$_2$ caused by **j**$_2$ is represented. In figure 4g **j**$_2$ is omitted and the translational movement is considered again. It leads to a new current **j**$_3$, figure 4h. In figure 4i **B**$_2$ and **v**$_{trans}$ are no longer shown but instead, the magnetic field **B**$_3$ generated by **j**$_3$. Next, the influence of the rotational movement is again taken into account, figures 4j and 4k: a current **j**$_4$ is generated. The last picture shows how the magnetic field **B**$_4$ is generated by **j**$_4$. The sequence of the cause-effect steps is summarized in Table 1.

$$\text{movement } \mathbf{v}_{trans} + \text{field } \mathbf{B}_0 \rightarrow \text{current } \mathbf{j}_1$$
$$\text{current } \mathbf{j}_1 \rightarrow \text{field } \mathbf{B}_1$$
$$\text{field } \mathbf{B}_1 + \text{movement } \mathbf{v}_{rot} \rightarrow \text{current } \mathbf{j}_2$$
$$\text{current } \mathbf{j}_2 \rightarrow \text{field } \mathbf{B}_2$$
$$\text{movement } \mathbf{v}_{trans} + \text{field } \mathbf{B}_2 \rightarrow \text{current } \mathbf{j}_3$$
$$\text{current } \mathbf{j}_3 \rightarrow \text{field } \mathbf{B}_3$$
$$\text{field } \mathbf{B}_3 + \text{movement } \mathbf{v}_{rot} \rightarrow \text{current } \mathbf{j}_4$$
$$\text{current } \mathbf{j}_4 \rightarrow \text{field } \mathbf{B}_4$$

Table 1. Four steps to get a field parallel to the starting field

We now have the desired result: $\mathbf{B}_4$ has the same direction as $\mathbf{B}_0$. So we no longer depend on the original field $\mathbf{B}_0$. Our "dynamo" continues to run, provided, of course, that the liquid moves fast enough.

The representation in Fig. 4 may have created a false impression. It looks as if we had "constructed" a stationary dynamo. The figure seems to show how a spiral movement reproduces the initial field $\mathbf{B}_0$. It is true that our model shows that such a field is generated. However, that does not mean that the field configuration is stationary. The reason is that we have ignored several other processes. In Figure 4d, for example, we only considered the rotation, but not the translation. The translational movement also generates currents by means of $\mathbf{B}_1$, and these, in turn, have their field, etc. In Figure 4g, we only considered the effect of the translation, but not that of the rotation. Again, we have disregarded other current components, etc.

Furthermore, we have not taken into account another effect that makes the field more complicated: We have assumed the flow to be given. We imposed it on the liquid. This is not true either. The magnetic fields have an effect on the liquid. The thermodynamic driving engine, i.e. convection, supplies the energy dissipated by the electric currents. It is thus subjected to a braking effect, like an ordinary technical dynamo to which a load is connected. This effect changes the flow of the liquid.

The actual process is,, therefore, more complex than Figure 4 suggests. More intricate structures are created and it cannot be expected that there is a stationary state. What we have looked at is only a small part of the overall picture. However, this contribution is important for an understanding of the geodynamo because we now see that the initial field $\mathbf{B}_0$ is no longer needed.

**IV. TIME AND LENGTH SCALES**

If an electrically conductive liquid is moved sufficiently irregularly electric currents and magnetic fields are generated. A simple experiment that shows this effect, one might think, looks like this: One fills a bucket with saltwater and stirs it, somewhat irregularly. From what has been said so far, we might expect electric currents to start flowing and magnetic fields to be created. But our common sense may also tell us that this will not happen. And in fact, it's not happening.

The reason why the experiment does not succeed is that the values of some of the parameters on which the effect depends are not large enough. In the following, we will ask what these parameters are.

**A. The lifetime of a current within the core of the Earth**

We leave aside the problem of the Earth's magnetic field for a moment and look at an *RL* circuit: the terminals of a coil are connected to the terminals of a resistor. We ask about the behavior of this arrangement when it is enlarged geometrically: when all linear dimensions are multiplied by one and the same factor $k$. We ask in particular how the decay time will scale. The original decay time being

$$\tau = \frac{L}{R}$$

we ask for the decay time

$$\tau' = \frac{L'}{R'}$$

after the circuit has been enlarged by a factor $k$. For this purpose, we need to calculate how the resistance and the inductance are scaling.

The resistance $R$ of a geometrically simple resistor can be calculated as

$$R = \rho \cdot \frac{l}{A}$$

Increasing the size of the resistor results in

$$R' = \rho \cdot \frac{l'}{A'}$$

where $l' = k \cdot l$ and $A' = k^2 \cdot A$.

Since the increased resistor consists of the same material as the original one, the specific resistance $\rho$ is not scaled. So we get

$R' = R/k$.

If the resistor is increased in this way by a factor of ten, the resistance $R$ is reduced to 1/10.

Accordingly, we calculate how the inductance $L$ scales. From the formula for the inductance

$$L = \mu_0 \cdot n^2 \cdot \frac{A}{l}$$

we get

$L' = k \cdot L$,

i.e. if a coil is enlarged by a factor of ten, the inductance increases tenfold.

With these results, we get for the decay time

$$\tau' = \frac{L'}{R'} = \frac{k \cdot L}{R/k} = k^2 \cdot \tau \tag{1}$$

The decay time of the $RL$ circuit thus increases with the square of the scaling factor. Let's look at an example: We assume an $RL$ circuit of laboratory size, with a linear dimension of about 0.1 m and a decay time of 1 millisecond. We don't even need a coil. Each closed circuit has an inductance and a resistance. We now imagine enlarging this circuit to 100 km, i.e. by a factor of $10^6$. Equation (1) tells us, that the decay time will increase to $10^9$ seconds or about 300 years. Closed currents on this scale thus remain practically constant for times of the order of 10 years. Significant changes are only observed in time intervals of the order of a hundred years.

We can transfer this result to the electric currents of the geodynamo. Each of the interdependent electric currents has a lifetime of decades to millennia, depending on the size of the corresponding current loop. We now understand why changes in the Earth's magnetic field can only be observed on large time scales. If the movement of the material of the Earth were suddenly stopped, the currents would continue to flow for many years to come, and the magnetic fields would continue to exist for such a time span.

## B. Conditions for a self-sustaining dynamo

In section III, we have seen how a dynamo effect can occur in a moving, electrically conductive liquid. However, there is one problem we have not yet addressed. The dynamo effect cannot occur if the newly created field is smaller than the initial one. This would mean that the dynamo would extinguish. To prevent this happening, some conditions must be met.

One of these conditions is that the individual currents do not die away too quickly. Their decay time must be large. As we have seen, the decay time of a circuit is greater the larger its geometric dimensions are. Moreover, $\tau$ is great when the liquid is a good electrical conductor.

Furthermore, the functioning of the dynamo depends on the velocity of the liquid. Just like a technical dynamo, a geodynamo can only work when the velocity of the moving electrical conductors is sufficiently high.

So we have identified three parameters on which the functionality of the dynamo depends:

(1) the geometric extension $l$, (2) the electrical conductivity $\sigma$ and (3) the velocity $v$ of the liquid.

These three conditions are related in the simplest way one can imagine: The product of $v$, $l$ and $\sigma$ must reach a certain minimum value. Multiplying this product by the magnetic field constant $\mu_0$ results in a dimensionless quantity, called the *magnetic Reynolds number*:

$R_m = \mu_0 \cdot \sigma \cdot v \cdot l$

The theoretical treatment of the problem shows that the minimum value for the occurrence of the dynamo effect is $R_m = 1$. However, for a geodynamo to run safely, $R_m$ must be larger.

We can now understand why the effect doesn't show up in the bucket with the saltwater.

With:

$\sigma = 1 \; \Omega^{-1} \cdot m^{-1}$
$v = 0.5 \; m/s$
$l = 0.2 \; m$

and $\mu_0 = 1.26 \cdot 10^{-6} \; Vs/(Am)$

we get

$R_m = 10^{-7}$,

a value that is far too small for a dynamo effect.

Somewhat more realistic would be the expectation that the movement of the water of the seas leads to a dynamo effect. Here a typical length would be about 100 m. But the magnetic Reynolds number is still far below the minimum value. So we shouldn't be surprised that we don't encounter this effect anywhere on Earth except in the real geodynamo.

This seems to show that it is hopeless to realize a geodynamo effect in a laboratory experiment. However, such experiments have proved successful[3]. The facilities are several meters in size and work with liquid sodium, which is pumped to a speed of up to 20 m/s. The path of the flowing sodium is predetermined. A magnetic Reynolds number of about 10 is achieved in these experiments.

We come back again to the explanation of the dynamo effect in section III. It seems that the complexity of the field strength distribution should increase with time. However, the smallest structures still have extensions of about 100 km. We can now understand this fact as well. Any smaller structure that would be created is not viable: The corresponding magnetic Reynolds number would be too small.

## V. THE INITIAL FIELD

In order for our dynamo to start, a magnetic field $\mathbf{B}_0$ at the beginning is needed. Two remarks can be made about the origin of this field.

1. If an electrically conductive liquid moves sufficiently quickly and irregularly, the field-free state is unstable. The system assumes the dynamo state "by itself". "By itself" means that a very small disturbance is sufficient to bring the system out of the field-free state. It is similar to a pencil that is placed on its tip. It tips over immediately, although there is a force-free state in which it would stand vertically. But hardly anyone will be surprised that it tips over anyway, and hardly anyone will ask why it falls in the direction in which it falls. Transferred to the Geodynamo, this means: The question about the starting field is not interesting. It is an example of spontaneous symmetry breaking.

2. Whoever is not satisfied with this statement must engage in investigating the mechanisms by which small electric currents, and thus magnetic fields, originate (similar to the person who wants to predict the direction in which the pencil tips, for example, has to deal with air currents and other small effects). Electrochemical or thermoelectric processes can be considered. For both, the conditions are given. However, these small causes, which were responsible for the starting of the dynamo, are certainly more difficult to investigate than the functioning of the running dynamo, because one asks for effects that were effective under the conditions that prevailed several billion years ago.

## VI. CONCLUSION

Although we are far from being able to calculate the Earth's magnetic field and its evolution over time, we now know the functional principles of the geodynamo and the reasons for its strange behavior. Let's summarize them once again.

For a dynamo effect to occur, an electrically conductive fluid must move helically.

The larger a circuit, the slower the current decays.

The occurrence of the dynamic effect depends on

- the velocity of the flow
- the electrical conductivity
- the geometric extension of the currents.

If the conditions for the occurrence of the dynamic effect are fulfilled, the currentless state is unstable.